# An electrically controlled single molecule spin switch

Wantong Huang[1*], Kwan Ho Au-Yeung[1*], Paul Greule[1], Máté Stark[1], Christoph Sürgers[1], Wolfgang Wernsdorfer[1,2], Roberto Robles[3], Nicolas Lorente[3,4], Philip Willke[1,5†]

[1] Physikalisches Institut, Karlsruhe Institute of Technology, Karlsruhe, Germany
[2] Institute for Quantum Materials and Technologies, Karlsruhe, Germany
[3] Centro de Física de Materiales CFM/MPC (CSIC-UPV/EHU), 20018 Donostia-San Sebastián, Spain
[4] Donostia International Physics Center, 20018 Donostia-San Sebastián, Spain
[5] Center for Integrated Quantum Science and Technology (IQST), Karlsruhe Institute of Technology, Karlsruhe, Germany

* These authors contributed equally to this work.
† Corresponding author: philip.willke@kit.edu

**Abstract**

Precise control of spin states and spin-spin interactions in atomic-scale magnetic structures is crucial for spin-based quantum technologies. A promising architecture is molecular spin systems, which offer chemical tunability and scalability for larger structures. An essential component, in addition to the qubits themselves, is switchable qubit-qubit interactions that can be individually addressed. In this study, we present an electrically controlled single-molecule spin switch based on a bistable complex adsorbed on an insulating magnesium oxide film. The complex, which consists of an Fe adatom coupled to an iron phthalocyanine (FePc) molecule, can be reversibly switched between two stable states using voltage pulses locally via the tip of a scanning tunnelling microscope (STM). Inelastic electron tunnelling spectroscopy (IETS) measurements and density functional theory (DFT) calculations reveal a distinct change between a paramagnetic and a non-magnetic spin configuration. Lastly, we demonstrate the functionality of this molecular spin switch by using it to modify the electron spin resonance (ESR) frequency of a nearby target FePc spin within a spin-spin coupled structure. Thus, we showcase how individual molecular machines can be utilized to create scalable and tunable quantum devices.

**Keywords**

Scanning Tunnelling Microscopy, Molecular machines, Spin states, Electron spin resonance, Transition metal complexes

**Introduction**

Individual electronic spins constitute promising building blocks for quantum information processing, quantum simulation, and quantum sensing. For that, several platforms are currently being explored which aim to harness spin qubit systems for quantum technologies including e.g. semiconductor quantum dots or color centers in diamond [1]. Another promising platform are molecular spin systems [2-5]: They are an appealing class, since they can be chemically designed and efficiently scaled up through self-assembly in long-range ordered structures. Towards the realization of functional quantum devices, not only the basic qubit units are required, but also a wide spectrum of functional units such as auxiliary qubits, electric control and tunable interqubit structures [5].

In this context, switchable qubit-qubit interactions are crucial for the implementation of multi-qubit gates: They allow one to rapidly control the interactions between individual qubits. These have been explored for molecular spins in both experiment and theory by a variety of approaches, including global MW pulses on different qubits [6,7] or on spin switches placed between them [8-10]. The latter approach, however, requires a local spin switch that can be altered either coherently or incoherently on a fast timescale. Moreover, these realizations still relied on ensemble electron spin resonance (ESR) experiments that do not grant access to a local control of individual qubits. In that regard, one realization was theoretically proposed, in which the tip of a scanning tunnelling microscope (STM) is used to implement two-qubit gates in polyoxometalate molecules: Here, two localized spins ($S = ½$) within the molecule can be coupled by injecting a tunnelling electron into the molecule's central core [11].

Single-molecule switches, as part of the framework of synthetic molecular systems coined artificial molecular machines [12], offer an alternative degree of freedom (mechanical, electronic or magnetic) as well as bistability that could potentially tune qubit-qubit interactions. They enable reversible transitions between stable states in response to external stimuli [12]: If integrated into spin circuits, they could control magnetic interactions between spin-containing molecules. For example, spin-crossover molecules [13] demonstrate switchable spin states via external stimuli such as electric fields, light or inelastic scattering with tunnelling electrons.

To probe molecular spins individually at the atomic scale, low-temperature STM is an excellent technique: It can precisely manipulate and characterize single molecules adsorbed on surfaces one-by-one. Recent experimental progress towards a quantum platform based on spins on surfaces has been made by combining ESR with STM [14,15]. For instance, this technique allowed for resolving magnetic coupling between atomic spins [16] and for a coherent control of single and multi-spin systems [17,18]. In addition, while the original work was based on individual atomic spins, an increasing number of molecular spin systems was investigated later on [19-24]. Also, magnetic switches in the form of rare earth atom stable magnets, Dy [25] and Ho [26], were realized that allowed to locally tune the resonance frequency of neighboring atomic spins, providing a foundation for exploring reversible spin qubit control. However, their stability remains limited by the onset of diffusion at elevated temperatures [27] and they require direct interaction with inelastic tunnelling electrons for the change of their magnetic state. These drawbacks make the realization of a spin switch in a molecular framework desirable. In return, while bistable molecular structures were established and investigated with STM on surfaces in great numbers [28-35], none was up to now shown to allow for tuning nearby magnetic states.

In this work, we investigate and employ a molecular spin switch that we can control electrically by the STM tip. We construct these bistable complexes, that consist of single Fe adatoms and FePc molecules, via tip-assisted on-surface assembly. We find that the complex is able to switch between two bistable states through the application of an STM bias voltage. DFT calculations of the two configurations highlight the crucial role of the Fe atom by changing the energetic landscape and thus enabling bistability. In combination with inelastic electron tunnelling spectroscopy (IETS) measurements and density functional theory (DFT) calculations, we identify a change in spin structure between the bistable states. Specifically, the spin S of the molecule reversibly switches between two states, S > 0 and S = 0. Utilizing its bistable spin states in electron spin resonance (ESR) experiments, we demonstrate the ability to alter the resonance frequency of a nearby FePc target spin center that is magnetically coupled to the spin switch.

**Results and discussion**

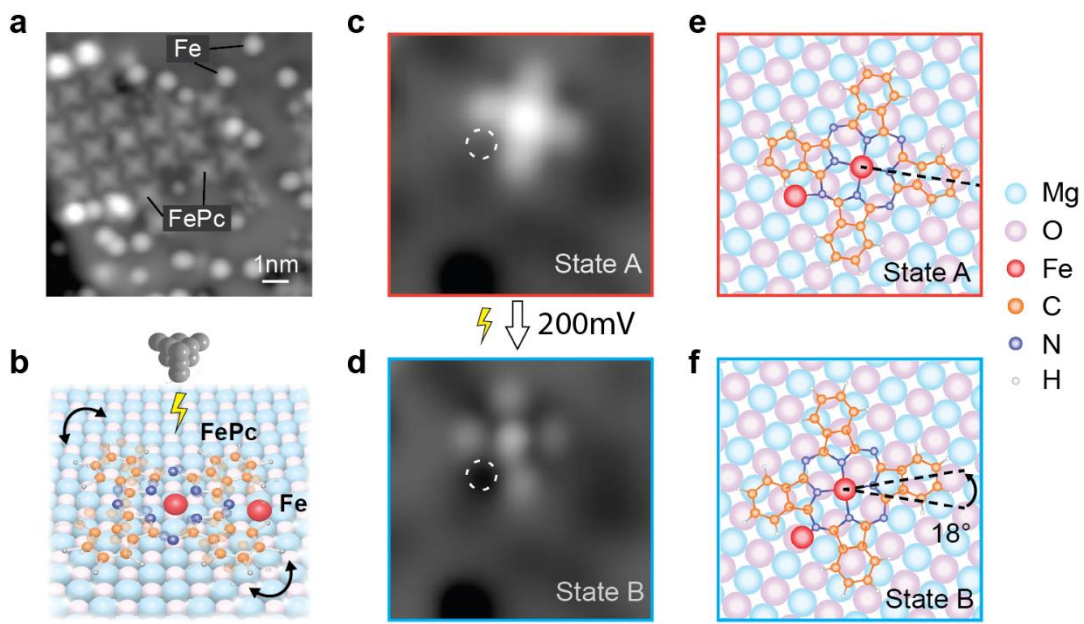

**Figure 1. Adsorption and bistability of Fe-FePc complexes on 2ML MgO/Ag(001).** (a) Overview STM image (10 nm x 10 nm; $V_{DC}$ = -200 mV, I = 60 pA) showing a self-assembled molecular island of pristine FePc molecules, and single Fe adatoms on 2ML MgO/Ag(001). (b) Schematic of a switchable Fe-FePc complex. STM images (3 nm x 3 nm) of the complex (c) before and (d) after applying a voltage pulse ($V_{DC}$ = 200 mV, I = 20 pA) over the molecule center. (I = 20 pA, c: $V_{DC}$ = -100 mV, d: $V_{DC}$ = 100 mV). The complex appears in a "bright" (State A) or "dark" (State B) contrast depending on its states. The Fe adatom position is indicated by a dashed circle. (e - f) Adsorption geometries (top view) obtained from DFT calculations of the Fe-FePc complex in (e) State A and (f) State B. In both cases, the Fe adatom is located in good approximation on an oxygen binding site of MgO, while FePc is located close to a Mg site in State B. The two configurations are rotated by 18 degrees.

The experiments were performed in a low-temperature STM with a base temperature of ~50 mK. Figure 1a shows a topographic image of the sample consisting of self-assembled molecular islands of pristine FePc along with individual Fe adatoms. FePc has been shown to be a mostly isotropic spin S = 1/2 system when adsorbed on MgO/Ag(001) [19, 20, 23], while individual Fe adatoms are a spin S = 2 with an out-of-plane magnetic anisotropy barrier $D = -4.6$ meV [36].

Previously, it was demonstrated that stable Fe-FePc complexes can be formed by STM tip manipulation: Here, the Fe adatom is located directly underneath one of the benzene ligands of FePc forming an Fe($C_6H_6$) half-sandwich complex that results in a reduction of the Fe atom spin state [23, 37]. In this study, we find that the same strategy, i.e. employing STM vertical manipulation, can be used to realize a bistable single molecule switch (Fig. 1b) when their relative alignment is slightly different: After picking up a single FePc and releasing it onto an Fe adatom on 2 ML MgO/Ag(001), the subsequent STM image (Fig. 1c) shows that the Fe-FePc complex has a similar cross-like appearance to the pristine FePc, but this complex (assigned as State A) exhibits a higher apparent height than the pristine one (Supplementary Fig. 1).

After applying an STM voltage pulse above the complex, the subsequent STM image shows that the Fe-FePc complex appears in a different contrast (Fig. 1d). The molecule shows clearly different features both in the center and on the ligands, and it has a lower overall apparent height of 70 pm compared to State A (Supplementary Fig. 1). Moreover, two of the four isoindole ligands are distorted, together with an additional dark contrast in between, indicating the position of the Fe adatom. We assign this Fe-FePc complex as State B, which is compared to State A rotated by about 18° around the Fe adatom. Based on a lattice site analysis (Supplementary Fig. 2) and the DFT-calculated adsorption geometries (Fig. 1e, f and Supplementary Fig. 3 and Supplementary Table 1), we conclude that, in both cases, the Fe adatom is situated between two isoindole ligands, but depending on the state, the alignment and lattice site is slightly different: for State A, both the Fe adatom and the FePc molecule center are adsorbed on an oxygen binding site of MgO, which is also the case for pristine Fe and FePc [19, 38]. In contrast, for State B, the FePc center is close to a Mg site of the MgO lattice. In both cases, the FePc center and Fe adatom are in close proximity: the MgO lattice difference between them is around (2, 0) for State A, and around (1.5, 0.5) for State B. Crucially, we find that the switching between the two states is highly controllable when applying bias voltage pulses (Supplementary Fig. 4).

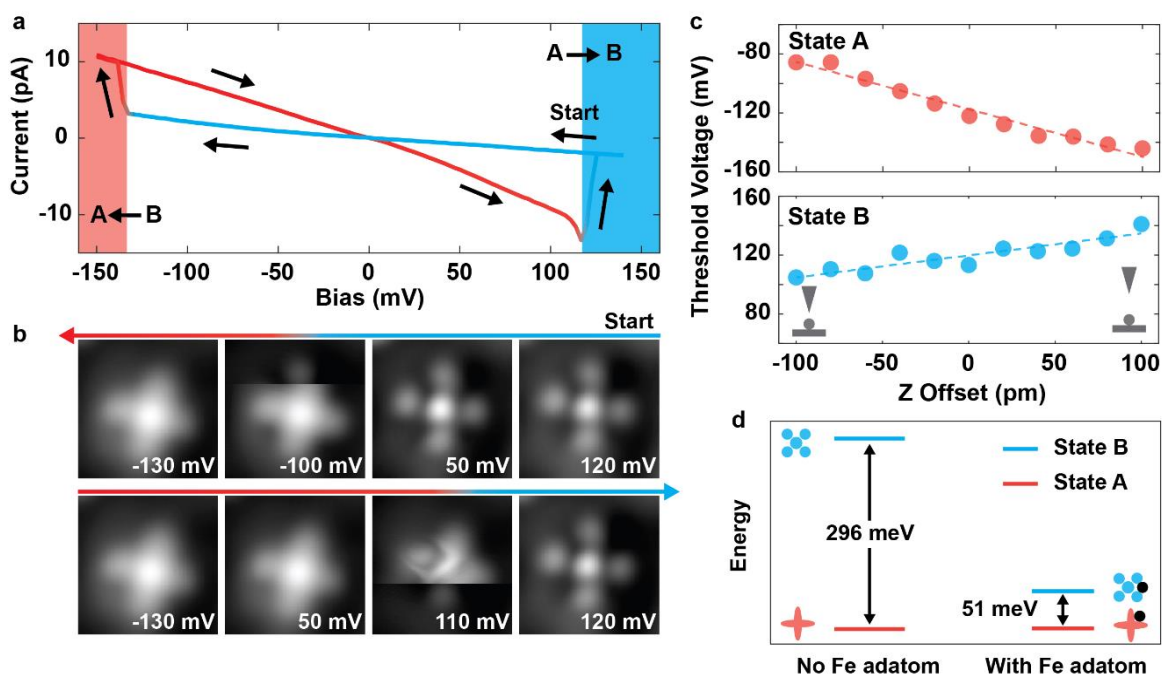

**Figure 2. STM-induced reversible switching of the Fe-FePc complex.** (a) The I(V) spectrum taken at the center of the Fe-FePc complex shows the switching between the two states, which additionally reveals a hysteresis behavior. Voltage sweep directions of decreasing and increasing bias are indicated by the arrows, starting from the positive side. The red and blue areas indicate the threshold voltage for switching between States A and B, respectively. (b) STM images (2 nm x 2 nm; I = 20 pA; scanning from top to bottom with forward and backward movement) taken at different biases that pinpoint the switching events. The colored arrows mark the scanning order. (c) Threshold voltage dependence on Z offset for State A (top) and State B (bottom). The feedback loop opens at zero z-offset ($V_{DC}$ = -130 mV, I = 20 pA, raw datasets are shown in Supplementary Fig. 8). Dashed lines represent linear fits, indicative of electric field-driven switching with slopes of -0.32 mV/pm for A, and 0.14 mV/pm for B, respectively. The sketches illustrate the proximity of the tip. (d) Comparison of absolute energies between State A and B with and without Fe adatom from DFT calculations. Without the Fe adatom, the energy difference is 296 meV, whereas with Fe, it is reduced to 51 meV.

In order to shed light on the switching mechanism between the two states, we perform I(V) spectra. Figure 2a shows a hysteresis behavior when sweeping the bias voltage across both polarities above the Fe-FePc complex. Together with the results of the bias-dependent STM images (Fig. 2b), it shows that the switching of the bistable conformation is bias polarity dependent. The switching threshold varies in the range of $V_{DC} = \pm(60 - 250)$ mV among different Fe-FePc complexes (Supplementary Fig. 5), possibly due to different tip geometries and local adsorption environments of the complex. In this specific case shown in Fig. 2a, the threshold voltage for B → A is at -133 mV and for A → B at +117 mV, respectively. For even higher bias voltages beyond the threshold voltage, we did not observe any switching between the two states. Instead, we observe a gradual tendency towards random rotation within a given state (Supplementary Fig. 6 and 7). In Fig. 2c, we plot the bias voltage as a function of tip height (Supplementary Fig. 8 for raw I(V) spectra), which reveals a linear relation $V_{thresh} \propto \Delta z$. This behavior would be in alignment with an electric field driven switching mechanism as found elsewhere for molecular systems on a surface [28, 39-43]. It also agrees with the observations of a nearly immediate response of the switching process, in contrast to what is expected for switching induced by rate-dependent inelastic tunneling events [44, 45]. Due to the small threshold voltages found here (~100 mV), we do not want to fully exclude the contribution from inelastic tunneling electrons for overcoming the potential energy barrier or a combination of both [46-49]. However, we note that the asymmetry in the stability with bias voltage is usually not expected from inelastic tunneling electron excitations. Moreover, we find that the switching is often also feasible when the tip is not located directly on top of the complex, but up to around a nanometer to the side. From the theoretical side, we find that, by employing DFT calculations, the difference in binding energy between State A and B is 51 meV (Fig. 2d and Supplementary Fig. 3). This shows that both states are very close in energy and supports our observation that they could easily be switched via the STM tip. In contrast, for pristine FePc, where the Fe adatom is absent, the energy difference between State A and B is 296 meV, and the controllable switch is not observed, suggesting that the Fe adatom is crucial for enabling the bistability.

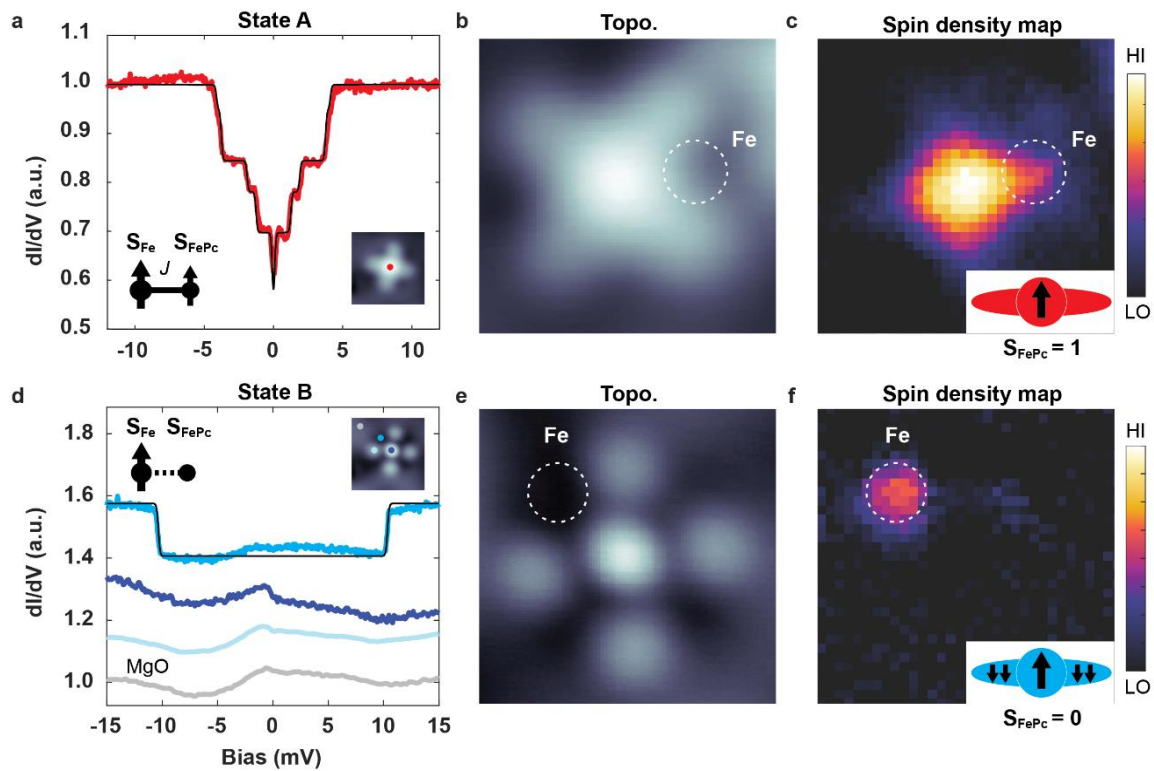

**Figure 3. Spin Structure of the Fe-FePc Complexes.** (a) dI/dV spectrum of State A acquired at the FePc center showing several steps originating from inelastic electron tunneling excitations [Step positions: ±(0.35, 1.30, 2.19, 3.90, 4.46) mV, setpoints: $V_{DC}$ = -30 mV, I = 100 pA, $V_{mod}$ = 50 μV]. The black curve is the fit using inelastic electron spin transport calculations [50], where an Fe spin S = 2 (D = -0.82 meV, E = 0.12 meV) is coupled to an FePc S = 1 (D = 1.29 meV), and the coupling strength is J = -0.85 meV. This is additionally illustrated in the sketch in the inset. (b) Topographic image of State A ($V_{DC}$ = -300 mV, I = 20 pA), and (c) The corresponding pseudocolor spin density map. (setpoint: $V_{DC}$ = -30 mV, I = 60 pA, $V_{mod}$ = 1 mV). The spin signal is mapped by the difference [dI/dV($V$ = 12 mV) - dI/dV($V$ = 0 mV)]. The Fe adatom position is indicated by a dashed circle. Additional dI/dV linecuts across the complex are shown in Supplementary Fig. 10. (d) dI/dV spectra of State B measured at different positions [from top to bottom: Fe adatom, FePc center, ligand and MgO (reference) as indicated in the inset ($V_{DC}$ = 30 mV, I = 80 pA, $V_{mod}$ = 0.3 mV)]. The step at ±10.5 mV is fitted by using $S_{Fe}$ = 2, D = -3.5 meV and $S_{FePc}$ = 0 (black curve). (e) STM image of State B ($V_{DC}$ = 80 mV, I = 20 pA)

and (f) the corresponding spin density map of the complex (setpoint: $V_{DC}$ = 30 mV, I = 80 pA, $V_{mod}$ = 0.8 mV). The spin signal is mapped by the difference [dI/dV(E = 15 meV) - dI/dV(E = 0 meV)]. Insets in (c) and (f) depict the sketch of the FePc spin states in State A and B, based on DFT calculations.

To probe the magnetic properties of our spin systems, we perform dI/dV spectroscopy measurements on the complex in both configurations. Figure 3a shows a dI/dV spectrum measured at the FePc center of the complex in State A, revealing several symmetric step features. We attribute these steps to inelastic electron tunneling spectroscopy (IETS) excitations from the magnetic ground state to the excited spin states. The magnetic nature of the IETS features is further supported by setpoint dependent dI/dV measurements with a magnetic tip (Supplementary Fig. 9). We additionally resolve the excitation by a spin contrast map in Fig. 3b, C emphasizing that the spin contrast is mostly concentrated on the center of the molecule and extends to a certain degree towards the ligands. The Fe adatom position is faintly visible as an enhanced IETS contrast between two of the lobes (indicated by a dashed circle). In contrast, the dI/dV spectra of State B show no spin signatures at either the FePc center or the ligand site. This is further supported in measurements on the FePc site using spin-polarized tips (Supplementary Fig. 9). However, a step feature appears at ~10.5 mV at the Fe adatom site, which is close to the inelastic excitation energy (~14 meV) of an isolated Fe adatom on MgO/Ag(001). The reduction in inelastic excitation energy likely arises from a different magnetic anisotropy. In the spin contrast map (Fig. 3f), this step feature is localized exclusively on the Fe adatom site (Fig. 3e). Thus, the measurements in Fig. 3 clearly demonstrate that the spin states are different between State A and B: State A shows a rich IETS pattern, suggesting that both spin centers host an unpaired spin and are likely coupled to each other. State B in contrast is spectroscopically silent on the FePc molecule, indicative of an $S = 0$ state. Only the Fe adatom shows inelastic excitations that are close to the features found for isolated Fe adatoms.

A similar change in the spin states is found in the DFT calculations (Fig. 1e, f; Supplementary Fig. 3). Table 1 summarizes the fractional charges found on the FePc center, the phthalocyanine ligand as well as the Fe adatom: According to the DFT calculations, the spin state of FePc in the complex transitions from a high-spin state ($S = 1$) in State A to a low-spin state ($S = 0$) in State B, while the Fe adatom retains a spin state of $S = 3/2$ in both configurations. In State A, FePc is antiferromagnetically coupled to the Fe adatom with an energy difference between ferromagnetic (FM) and antiferromagnetic (AFM) coupling ($E_{FM} - E_{AFM} = 4$ meV). In contrast, in State B, both the central Fe atom of FePc and the phthalocyanine ligand host a spin $S = 1$. Since both are preferentially antiferromagnetically coupled, they form a non-magnetic configuration, resulting in $S = 0$ configuration. Moreover, the central Fe atom of FePc and the Fe adatom exhibit a preference for ferromagnetic coupling ($E_{FM} - E_{AFM} = -45$meV). Thus, the change in spin configuration of FePc is rationalized by a change in the electronic occupation of both the central Fe spin as well as its ligand spin, while the Fe adatom maintains an unchanged spin state in both cases.

|  | FePc (Fe) | FePc (Pc) | FePc (total) | Fe adatom |
|---|---|---|---|---|
| State A (DFT) | -2.09 $\mu_B$ (S=1) | 0.01 $\mu_B$ (S=0) | -2.08 $\mu_B$ (S=1) | 3.07 $\mu_B$ (S=3/2) |
| State A (IETS) |  |  | S = 1 | S = 2 |
| State B (DFT) | 1.84 $\mu_B$ (S=1) | -1.69 $\mu_B$ (S=1) | 0.16 $\mu_B$ (S=0) | 2.91 $\mu_B$ (S=3/2) |
| State B (IETS) |  |  | S = 0 | S = 2 |

**Table 1. Spin configuration in the two states found in experiment and DFT.** The sign of the spin in DFT indicates the relative alignment between different spins. We find that the main difference between the two states is the additional occupation of the ligand spin in the case of State B.

To gain also a deeper understanding of the spin structure from the experiment, we simulate the dI/dV spectra for both configurations using spin transport calculations [50]. For the complex in State A, both the correct position and intensity of IETS measurements can be best reproduced (black curves in Fig. 3a) using a Hamiltonian of the form

$$H = J\vec{S}_{FePc} \cdot \vec{S}_{Fe} + D_{FePc}S^2_{FePc,z} + \left[D_{Fe}S^2_{Fe,z} + E_{Fe}\left(S^2_{Fe,x} - S^2_{Fe,y}\right)\right] \quad (1)$$

where $J = -0.85$ meV (FM) is the Heisenberg exchange coupling between the FePc and Fe adatom spins. In this simulation, FePc has a spin state of $S_{FePc} = 1$ with an out-of-plane magnetic anisotropy $D_{FePc} = 1.29$ meV. The Fe adatom has a spin state of $S_{Fe} = 2$ with an out-of-plane magnetic anisotropy $D_{Fe} = -0.82$ meV and an in-plane magnetic anisotropy $E_{Fe} = 0.12$ meV (See Supplementary Fig. 11 for the energy level diagram). For the complex in State B, the observed step feature at the Fe adatom is reproduced with $S_{Fe} = 2$ and $D_{Fe} = -3.5$ meV (Fig. 3d black curve) while the spin of FePc is set to $S_{FePc} = 0$, thus being non-magnetic. Although the spin state of the Fe adatom differs here from the DFT calculation, we suggest that it remains $S_{Fe} = 2$, since the IETS result closely resembles that of an isolated Fe adatom ($S_{Fe} = 2$) [36]. Also, spin models using $S_{Fe} = 3/2$ were tested and failed to reproduce the data for State A. In general, a great variety of other spin configurations were tested, of which the chosen set of spin states and parameters $J$, $D$ and $E$ provided the best agreement with the experimental spectra. In addition to the difference in Fe spin state, the spin coupling in State A is AFM in the DFT calculations, while experimentally we find FM coupling. We rationalize this by the fact that magnetic exchange couplings are generally difficult to capture by DFT and that the coupling is overall rather weak for State A (DFT: $E_{FM} - E_{AFM} = 4$ meV; IETS: $J = -0.85$ meV). More importantly, we highlight that both DFT and IETS find a change from a paramagnetic (State A, $S_{FePc} = 1$) to a non-magnetic configuration (State B, $S_{FePc} = 0$) of the FePc molecule driven mainly by the change in spin occupation of the Pc ligand.

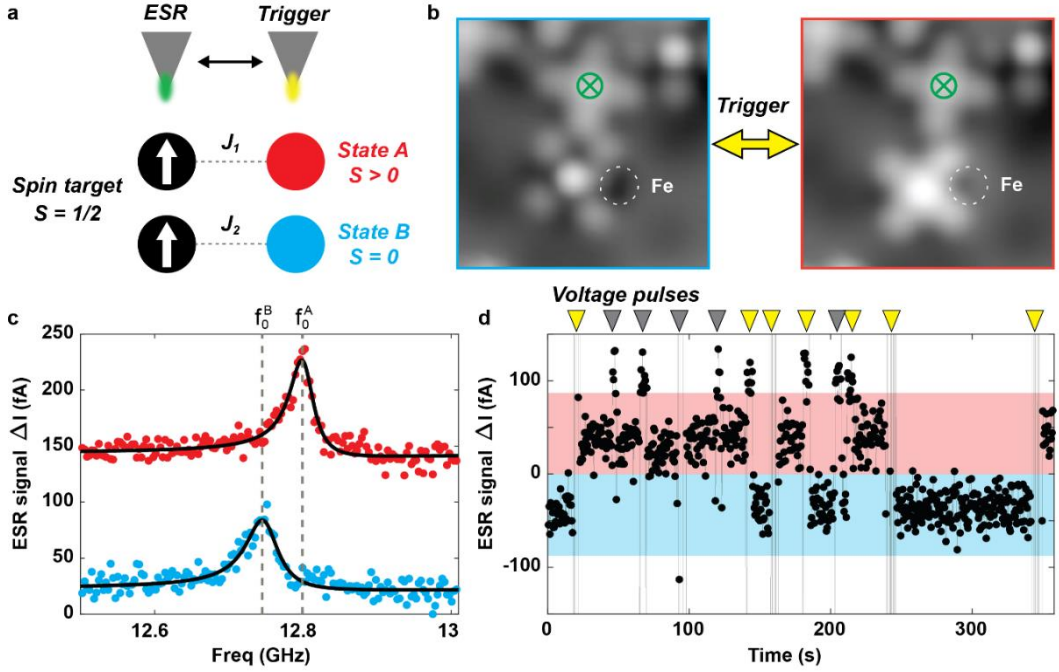

**Figure 4. Electron spin resonance (ESR) frequency shift using the spin switch.** (a) Schematic of the experimental setup: an FePc spin ½ target is coupled to an Fe-FePc spin switch. The distance between the molecule centers is (3, 5) MgO lattice sites (Supplementary Fig. 12). The tip is used for ESR readout when positioned above the target spin. For triggering the switching events it is moved above the spin switch. (b) STM images (3 nm x 3 nm; $V_{DC}$ = -100 mV, I = 20 pA) showing an Fe-FePc spin switch (bottom) coupled to a target FePc molecule (top, green cross) at the edge of a molecular island. Switching events were induced by an STM voltage pulse (State B → A: $V_{DC}$ = -300 mV, I = 20 pA; State A → B: $V_{DC}$ = 300 mV, I = 500 pA). (c) ESR signals recorded at the center of the FePc target spin, showing a 53 MHz shift between the resonance frequencies $f_0^A$ and $f_0^B$ of State A (top) and State B (bottom), respectively. (ESR conditions: $V_{DC}$ = 40 mV, I = 10 pA, $V_{rf}$ = 8 mV, $B_{ext}$ = 484 mT). (d) ESR signal $\Delta I$ with fixed frequency set to $f_{\text{probe}} = f_0^A$ recorded over time. A high signal (red region) corresponds to on-resonance conditions of $f_0^A$, while a low signal (blue region) corresponds to off-resonance conditions of $f_0^A$. Triangles at the upper end of the graph indicate switching protocols, in which the tip is briefly moved to trigger the spin switch via corresponding voltage pulses (State B → A: $V_{DC}$ = -300 mV, State A → B: $V_{DC}$ = 300 mV) and return to the target spin as illustrated in a. In total 12 voltage pulses were attempted to induce the switching event, of which 7 were successful (yellow). Spikes in the current are due to the tip movement and change of the tunneling parameters for switching during the ESR measurement (ESR conditions: $V_{DC}$ = 40 mV, I = 10 pA, $V_{rf}$ = 8 mV, $B_{ext}$ = 508 mT, $f_{probe} = 13.74$ GHz, tip travel speed = 18.4 nm/s. Note that the different frequency compared to (c) results from a different external magnetic field. All the voltage pulses and ESR measurements were done with closed feedback loop).

In order to showcase the operation of this spin switch, we built a small structure that allows us to test its action on a target spin system. The control mechanism is outlined in Fig. 4a while the structure is shown in Fig. 4b. It consists of a pristine FePc (Fig. 4b: top) acting as a simple S = ½ target spin system with $g \approx 2$ [19, 20] as well as the bistable Fe-FePc switch (Fig. 4b: bottom) in direct proximity. The resonance frequency $f_0$ of the FePc is given by

$$hf_0 = g\mu_B B \qquad (2)$$

Where $B$ is the external magnetic field, $h$ is Planck's constant and $\mu_B$ the Bohr magneton. However, $f_0$ can be additionally shifted by neighbouring magnetically coupled spins [16, 19, 26]. This adds a term $hf_{dipolar} = \frac{\mu_0}{2\pi}\frac{1}{r^3}m_z^{FePc} \cdot m_z^{switch}$ to Eq. (2), that can shift the resonance frequency of the target spin as well, depending on whether the spin switch is $S = 0$ or $S = 1$. In a similar fashion, Heisenberg exchange interaction, $H = J \cdot S_{switch} \cdot S_{FePc}$, can shift $f_0$ [20]. To test the influence of the spin switch, we measured an ESR spectrum [15, 19] (see method section) on the target FePc (Fig. 4c). With the switch in State B, we obtain a resonance frequency $f_0^B = 12.749\ \text{GHz}$, which corresponds to good approximation to that of a spin ½ with $g \approx 2$. Subsequently, we applied a voltage pulse to switch to State A (Fig. 4b: right). Now, using the same ESR parameters, we obtain $f_0^A = 12.802\ \text{GHz}$ as shown in Fig. 4c. In other words, the switching between State A and B of the close-by spin switch results in a frequency shift of $\Delta f_0 = 53\ \text{MHz}\ (\approx 220\ \text{neV})$ of the target FePc molecule. This shift is larger than expected for bare dipolar interaction (22 MHz), as estimated using the spin states from IETS results. We attribute the additional shift to ligand-mediated ferromagnetic exchange interaction between the two spins [19] (See Supplementary Text for discussion on the spin-spin coupling).

To demonstrate a reversible switching between the two states, we established a switching routine in Fig. 4d. Here, we continuously monitor the ESR signal over time at the resonance condition of State A, i.e. $f_{\text{probe}} = f_0^A$. Consequently, a high ESR signal probes the ESR peak under resonance conditions (red range), while a low signal indicates that $f_0^A$ is shifted away from $f_{\text{probe}}$ (blue range). Next, we toggled the resonance condition between States A and B by a spin switch routine: This routine moved the tip from the target FePc to the spin switch and applied an STM voltage pulse for 3 s. After triggering the switch, the tip was immediately returned to the target FePc position for continued ESR readout. In total, we performed 12 switching attempts, 7 of which were successful (yellow triangles). The success rate here is not as high as that of an isolated spin switch (Fig. 2 and Fig. 3), likely due to the presence of the nearby target FePc. The success rate could be further improved by fine tuning of the switching parameters (In particular for B→A, which accounts for 4 of the 5 failed attempts) and by simply applying multiple pulses at the same time. In addition, we found that only switching from State B to A actually required additional lateral movement of the tip: The switching from State A to B remained even feasible remotely, i.e. with the tip kept at the target FePc position.

B to A actually required additional lateral movement of the tip: The switching from State A to B remained even feasible remotely, i.e. with the tip kept at the target FePc position.

**Conclusion**

In summary, we have experimentally demonstrated a simple atomic structure, consisting of an Fe adatom and an FePc molecule, that shows bistability in its ground state configuration. This bistable switch shows a distinct change in its magnetic structure. In particular, we were able to demonstrate its functionality by altering the resonance frequency of a target spin system. Our work prototypes the construction of a scalable switchable spin architecture in terms of single molecule machines. It highlights how the local energy landscape of on-surface objects such as atoms and molecules can be exploited by electric control of the STM tip to alter spin systems. The presented switch is applicable in on-surface spin qubit structures; however, we stress that the underlying operational principles may also be achieved through the chemical synthesis of specialized spin switch molecules: First, the addition of the Fe atom brings the two states closer in energy enabling the electric field-induced bistability. Thus, the Fe adatom and the surface play a crucial role here in promoting bistability, which instead could be realized through the incorporation of specific side groups in a single spin switch molecule. Second, the magnetic functionality of the switch is enabled by a change in the molecule's spin occupation loading a spin to its ligand, thereby compensating the total spin.

Thus, we envision the implementation of different classes of single molecule machines with the general ingredients presented here, that are essential to develop molecule-based spintronics and quantum information devices.

**Methods**

The sample preparation was carried out *in-situ* at a base pressure of $< 5 \times 10^{-10}$ mbar. The Ag(001) surface was prepared through several cycles of Argon ion sputtering and annealing through e-beam heating. For MgO growth, the sample was heated up to 430° C and exposed to a Mg flux for 20 minutes in an oxygen environment at $10^{-6}$ mbar leading to a MgO coverage of ~50% and layer thicknesses ranging from 2 to 5 monolayers. Subsequently, FePc was evaporated onto the sample held at room temperature using a home-built Knudsen cell at a pressure of $9 \times 10^{-10}$ mbar for 90 seconds. Electron-beam evaporation of Fe was carried out for 21 seconds onto the cold sample. We determined the thickness of MgO layers through point-contact measurements on single Fe adatoms [51]. All experiments were carried out using a Unisoku USM1600 STM inside a homebuilt dilution refrigerator with a base temperature of 50 mK. An effective spin temperature of ~300 mK was estimated from ESR measurements of Fe dimers. Here, the intensities of the electron spin states depend on temperature [16], which we take as an estimate of the Boltzmann distribution in the experiment.

STM vertical manipulation was employed to build Fe-FePc complexes. Firstly, by positioning the tip above one of the ligands of the FePc molecule on MgO/Ag(001), the molecule is picked up by gently approaching the tip close to the molecule. Next, we apply a short STM voltage pulse at V = 0.85 V with an opened feedback loop. A sudden change of the tunneling signal can be observed, indicating a successful pick up of molecule. Then, a subsequent topography is recorded to ensure the pick-up. After that, the tip will be positioned above an Fe adatom on MgO/Ag(001) at a predefined position. Lastly, a similar sequence of "pick up" is then applied to drop the molecule onto the Fe adatom.

Spin-polarized tips were prepared as follow: (1) individual Fe adatoms were transferred onto the Ag coated PtIr tip by STM vertical manipulation. (2) The spin polarization was then verified through the asymmetry in the differential conductance around zero bias in (dI/dV) measurements on FePc adsorbed on MgO/Ag(001). (3) Magnetic tips showing a high spin contrast were subsequently tested in the ESR-STM measurements (continuous wave). The radiofrequency (RF) voltage was applied on the tip-side of the junction using a RF generator (Rohde & Schwarz SMB100B). The RF voltage was combined with the DC tunnel bias using a Bias tee (Marki Microwave MDPX-0305). We used a digital lock-in amplifier (Stanford Research Systems SR860) to read out the ESR signal using an on/off modulation scheme at 323 Hz. Note that while the bias voltage was applied to the STM tip, all bias signs were inverted in the manuscript to follow the conventional definition of bias voltage with respect to the sample bias.

DFT calculations were performed using the VASP code [52]. The PBE form of the GGA exchange-correlation functional was used [53], and missing dispersion interactions in this functional were treated using the D3 scheme with Becke-Johnson damping [54]. The core electrons were treated by the projector augmented-wave method [55], and wave-functions were expanded using a plane wave basis set with an energy cutoff of 400 eV. The Dudarev implementation of the LDA+U method [56] was used to treat the 3$d$ electrons of Fe, with $U_{eff}$ = U-J = 3 eV, which has been used in previous FePc studies [57, 58]. The MgO/Ag(001) surface was modeled using a slab formed by two MgO layers on top of four Ag layers, with a vacuum region of at least 15 Å, and a 6 × 6 surface unit cell. The position of all atoms in the unit cell except the two bottom Ag layers were relaxed until forces were smaller than 0.01 eV/Å. Corrections to potential and forces due to the presence of a dipole moment in the slab were applied [59]. Charge transfers and magnetic moments were determined by Bader analyses [60].

## Acknowledgements

P.W. acknowledges funding from the Emmy Noether Programme of the DFG (WI5486/1-1), financing from the Baden Württemberg Foundation Program on Quantum Technologies (Project AModiQuS). P.W. and K.H.A.Y. acknowledge support and financing from the Centre for Integrated Quantum Science and Technology (IQST). P.G. and P.W. acknowledges financial support from the Hector Fellow Academy (Grant No. 700001123). R.R. and N.L. thank financial support from project PID2021-127917NB-I00 funded by MCIN/AEI/10.13039/501100011033, from project IT-1527-22 funded by the Basque Government, and from project ESiM no. 101046364 funded by the European Union.

## Notes

The authors declare no competing financial interests.